\documentclass[aps,pra,twocolumn,superscriptaddress,longbibliography]{revtex4-2}
\usepackage{graphicx}
\usepackage{latexsym}
\usepackage{amssymb}
\usepackage{amsmath}
\usepackage{amsfonts}
\usepackage{upgreek}
\usepackage{float}
\usepackage{bm}
\usepackage{multirow}
\usepackage{color}
\usepackage[T1]{fontenc}
\usepackage{hyperref}
\usepackage{subfigure}
\usepackage{booktabs,tabularx}

\hypersetup{
colorlinks = true,
linkcolor = [rgb]{0.70,0.13,0.13},
citecolor = [rgb]{0.13,0.55,0.13},
urlcolor  = [rgb]{0.25, 0.41, 0.88}}
\newcommand{\ket}[1]{|#1\rangle}
\newcommand{\bra}[1]{\langle#1|}

\begin{document}
\title{Many-body entanglement and spectral clusters in the extended hard-core bosonic Hatano-Nelson model}

\author{Chao-Ze Lu}
\affiliation{College of Physics, Nanjing University of Aeronautics and Astronautics, Nanjing, 211106, China}
\affiliation{Key Laboratory of Aerospace Information Materials and Physics (NUAA), MIIT, Nanjing 211106, China}

\author{Gaoyong Sun}
\thanks{Corresponding author: gysun@nuaa.edu.cn}
\affiliation{College of Physics, Nanjing University of Aeronautics and Astronautics, Nanjing, 211106, China}
\affiliation{Key Laboratory of Aerospace Information Materials and Physics (NUAA), MIIT, Nanjing 211106, China}

\begin{abstract}
We study many-body entanglements and spectra of the extended bosonic Hatano-Nelson model in the hard-core limit. 
We show that the system undergoes a phase transition from a gapless phase to a charge density wave phase accompanied by a $\mathcal{PT}$ transition in the first excited state.
The phase transition is characterized by the crossing of the ground-state biorthogonal order parameter and the sudden change of the first excited-state entanglement entropy.
The gapless phase is verified by the logarithmic scaling of the ground-state entanglement entropy with the central charge $c=1$.
Furthermore, we show that all energy spectral clusters would form ellipses in strong nearest-neighbor interactions, for which we establish a universal scaling law. 
The lengths of the major and minor axes are shown to obey power laws with respect to the nearest-neighbor interaction.
The exact expressions are derived for the numbers of energy levels on the outermost elliptic ring of each clusters.

\end{abstract}

\maketitle

\section{Introduction}
Quantum entanglement that is a central concept of quantum mechanics has broad potential applications, 
such as quantum teleportation \cite{bouwmeester1997experimental}, quantum computation \cite{nielsen2010quantum, jozsa2003role} and quantum sensing \cite{degen2017quantum}
in quantum technology and quantum information science \cite{nielsen2010quantum, horodecki2009quantum}.
A notable application of the quantum entanglement for a many-body system is to characterize the equilibrium phase transition \cite{osterloh2002scaling, eisert2010colloquium},
where the entanglement entropy was shown to display a logarithmic divergence in a one dimensional model \cite{vidal2003entanglement}.
Quantum entanglement is a crucial tool in understanding quantum phases of Hermitian many-body systems.

Non-Hermitian systems are of particular interest because of many unique phenomena that have no counterparts in Hermitian systems \cite{bergholtz2021exceptional,ashida2021non}.
It is known that non-Hermitian skin effects \cite{lee2016anomalous,yao2018edge,kunst2018biorthogonal,xiong2018does,
gong2018topological,alvarez2018non,yokomizo2019non,okuma2020topological,zhang2020correspondence,yang2020non,
wang2020defective,jiang2020topological,weidemann2020topological,xiao2020non,borgnia2020non} 
and exceptional points \cite{heiss2012physics,kozii2017non,hodaei2017enhanced,zhou2018observation,miri2019exceptional,park2019observation,yang2019non,
ozdemir2019parity,dora2019kibble,jin2020hybrid,xiao2021observation,chen2022asymmetric}
are two of fascinating phenomena in non-Hermitian systems.
Recently, the entanglement entropy has been generalized to non-Hermitian systems for understanding phase transitions.
For instance, the biorthogonal entanglement entropy is introduced to describe the equilibrium phase transitions \cite{chang2020entanglement,guo2021entanglement,chen2022quantum,sanno2022engineering,tu2022renyi,bianchini2014entanglement,herviou2019entanglement}, 
where the central charge is found to negative at an exceptional point from the entanglement entropy under specific treatments\cite{chang2020entanglement,sanno2022engineering,tu2022renyi}. 
The negative central charge is argued to be described by the nonunitary conformal field theory \cite{chang2020entanglement,sanno2022engineering}. 
The understanding of the entanglement entropy at an exceptional point remains an open question.

Many-body physics of non-Hermitain systems is another interesting topic, in which rich unique phenomena
such as many-body skin effects \cite{zhang2020skin,mu2020emergent,lee2021many,kawabata2022many,zhang2022symmetry,alsallom2022fate,shen2022non,sun2023aufbau}, 
many-body edge bursts \cite{hu2023many}
and entanglement phase transitions \cite{kawabata2023entanglement,feng2023absence,li2023emergent,li2023entanglement}
are explored in recent years. 
Concerns for many-body spectra \cite{kawabata2022many,zhang2022symmetry,sun2023aufbau} might be at the core of the researches as a direct extension of the single-particle physics.
Recently, studies show that interactions can induce $\mathcal{PT}$ transitions \cite{zhang2022symmetry} and spectral structures \cite{kawabata2022many,zhang2022symmetry} in the interacting fermionic Hatano-Nelson (FHN) model under periodic boundary conditions (PBCs).

\begin{figure}[tb]
\includegraphics[width=8.6cm]{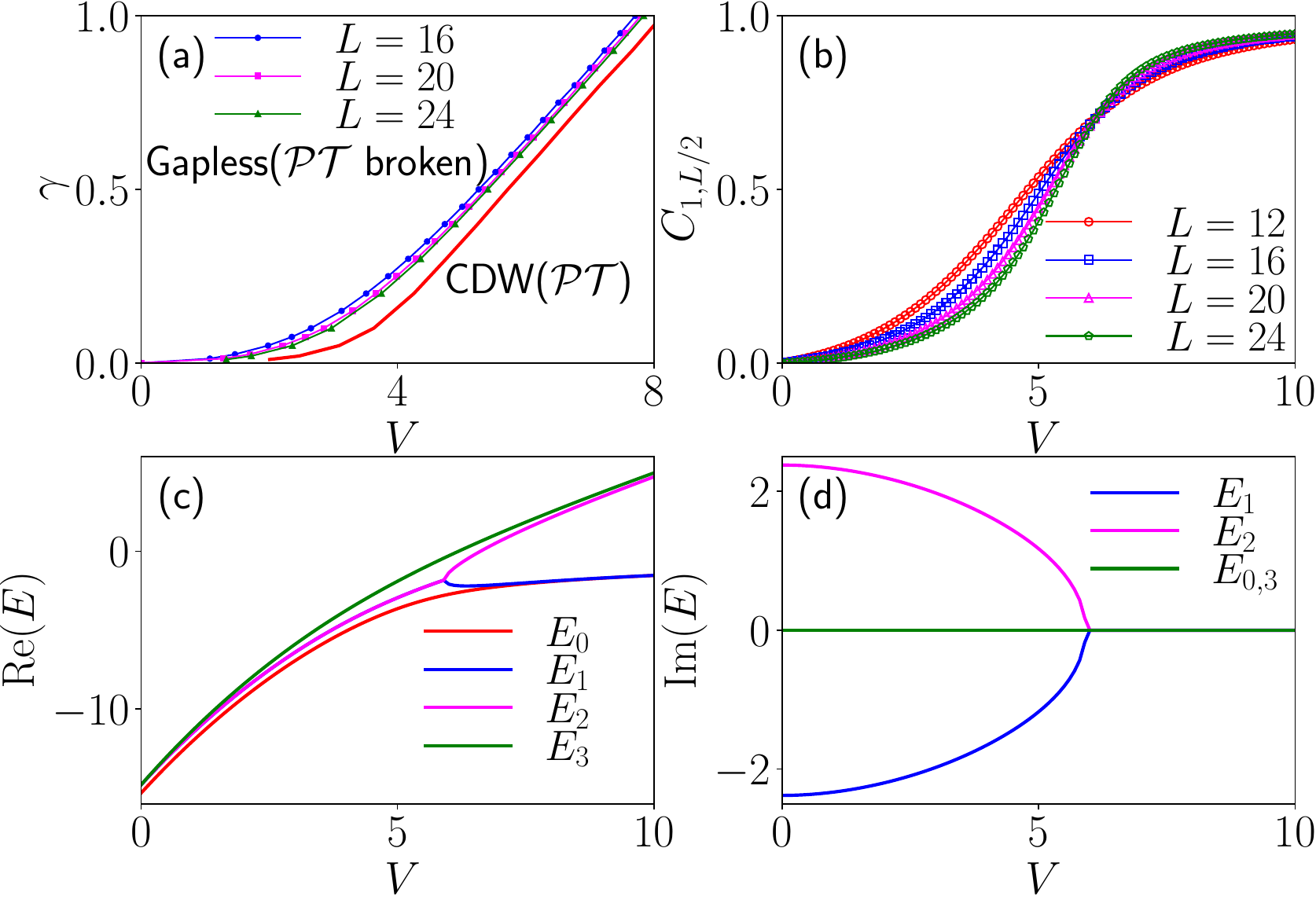}
\caption{ Many-body phase diagram of the half-filled hard-core BHN model with respect to $\gamma$ and $V$ at $t=1$ up to $L=24$ lattice sites in PBCs,
(a) The phase diagram obtained from the change of the first-excited energy as shown in (c) and (d), in which the red solid line denotes critical points in the thermodynamic limit derived by extrapolations.
(b) The biorthogonal correlation function $C_{1,L/2}$ at $\gamma=0.6$, where the crossing point denotes the critical point.
(c) The four lowest real parts of eigenenergies at $\gamma=0.6$.
(d) The imaginary parts of four eigenenergies shown in (c).} 
\label{figPhaseDiagram}
\end{figure}

In order to further understand phase transitions and spectral structures induced by the interaction, we instead study the extended interacting bosonic Hatano-Nelson (BHN) model with a hard-core constraint, which is inequivalent to FHN under PBCs even in the thermodynamic limit \cite{sun2023aufbau}.
We show that the BHN model undergoes a phase transition in the ground state accompanied by a parity-time ($\mathcal{PT}$) transition in the first excited state. 
The ground-state phase transition can be described by the biorthogonal order parameter. 
The central charge $c=1$ for the gapless phase is derived from both the biorthogonal entanglement entropy and the self-norm entanglement entropy.
For the first-excited phase transition, we find that the biorthogonal and self-norm entanglement entropy display a sudden change near the critical point. 
In addition, we study the many-body energy spectra by increasing the interaction, where we find that the energy spectral clusters form ellipses in strong nearest-neighbor interactions.
We establish the universal laws of spectral clusters, and show that the structure of an ellipse dependents on the interaction, the filling and the number of clusters.

This paper is organized as follows. 
In Sec.\ref{sec:HNModel}, we introduce the bosonic Hatano-Nelson model.
In Sec.\ref{sec:Entanglement}, we study the entanglement entropy of the ground state and the first excited state.
In Sec.\ref{sec:EnergySpectrum}, we discuss the properties of spectral clusters in strong nearest-neighbor interactions.
In Sec.\ref{sec:conclusion}, we summarize our results.

\section{Model}
\label{sec:HNModel}
In this paper, we study the one-dimensional extended BHN model, which is a non-Hermitian system with a nearest-neighbor nonreciprocal hopping. The Hamiltonian of the extended BHN model can be written as,
\begin{equation}
	H=\sum_{l=1}^{L} [ (t + \gamma)b_l^{\dagger}b_{l+1} + (t - \gamma)b_{l+1}^{\dagger}b_l
	+ \frac{U}{2} n_l(n_{l}-1) + V n_ln_{l+1} ],
	\label{eqHN}
\end{equation}
where $b_l^{\dagger} (b_l)$ is the creation (annihilation) operator of a boson at the lattice site $l$, and $n_l=b_l^{\dagger}b_l$ is the bosonic number operator. $L$ is the length of the chain.
The real parameters $ t $ and $\gamma$ denote the reciprocal and nonreciprocal hopping coefficients of bosons between two nearest neighboring sites, respectively. The coupling coefficient $U$ is the on-site interaction, and $V$ is the nearest-neighbor interaction between two bosons in adjacent lattice sites.
The PBCs is imposed by $b_{L+1}=b_{1}$.

Because it is extremely difficult to diagonalize a huge non-Hermitian matrix, numerical simulations of soft-core bosons is beyond the scope of the exact diagonalization method. In the following, we will study the system with the exact diagonalization in PBCs and merely consider the system with a hard-core constraint ($U \rightarrow \infty$).
The extended BHN model in the hard-core limit is given by,
\begin{equation}
	H=\sum_l [ (t + \gamma)b_l^{\dagger}b_{l+1} + (t - \gamma)b_{l+1}^{\dagger}b_l + V n_ln_{l+1} ],
	\label{eqHNHC}
\end{equation}
with a constraint $n_l=\{0,1\}$.
The BHN model in the hard-core limit is equivalent to the FHN model in OBCs \cite{zhang2022symmetry}, but is different from the FHN model in PBCs \cite{sun2023aufbau}.
The eigenvalues of the BHN model in Eq.(\ref{eqHNHC}) are always real for $\gamma < t$ in OBCs as the Hamiltonian can be mapped to a Hermitian Hamiltonian under a site-dependent similarity transform \cite{sun2023aufbau}. 
In the following, we will set $t=1$ during the simulations and study the extended hard-core BHN model in Eq.(\ref{eqHNHC}) under PBCs as we are interested in the properties of the complex-valued energy spectra.

\begin{figure}[tb]
\includegraphics[width=8.6cm]{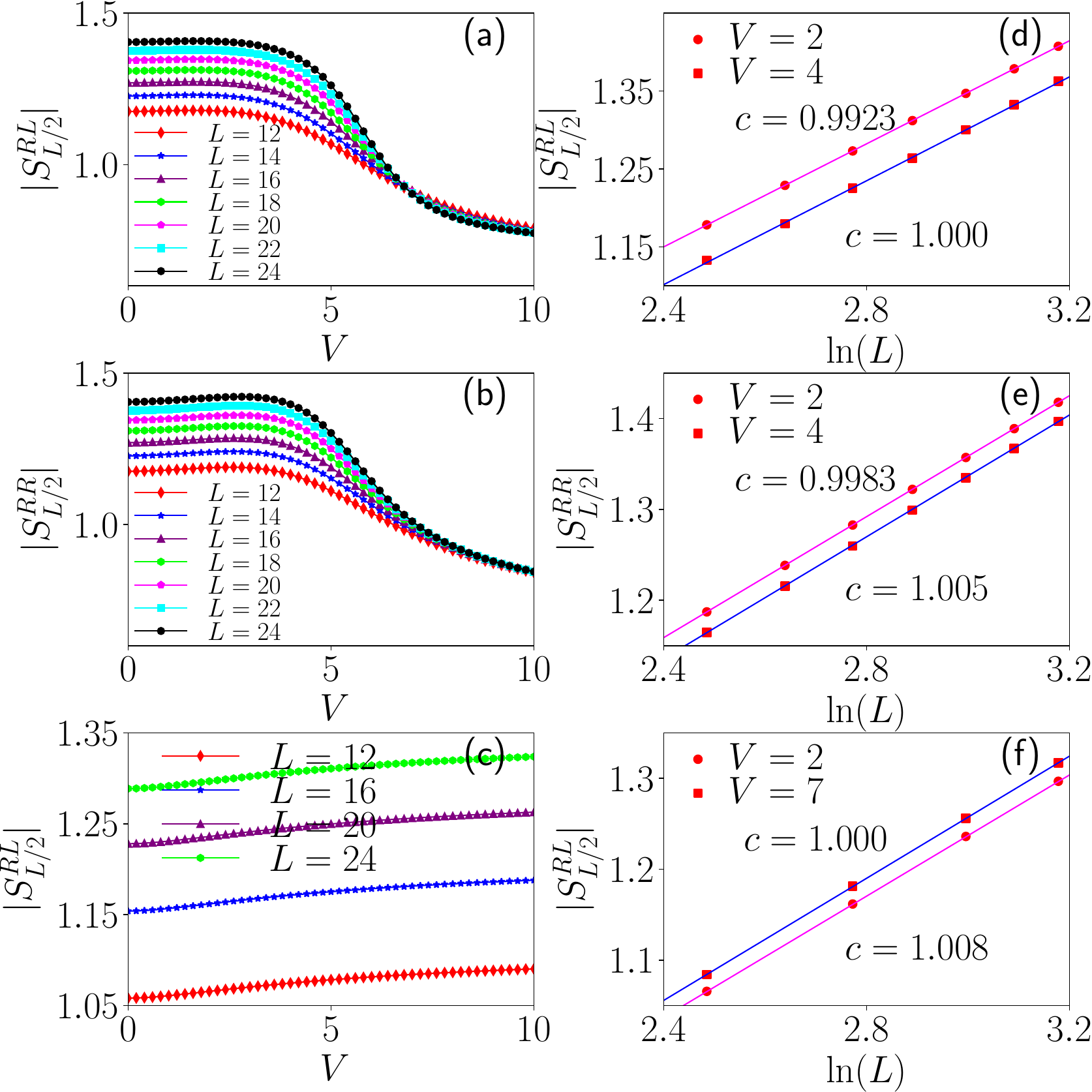}
\caption{ Half-chain entanglement entropy of the BHN model at $t=1$ and $\gamma=0.6$ in PBCs.
(a)(b)(c) The biorthogonal entanglement entropy in the half filling, the self-norm entanglement entropy in the half filling,
and the biorthogonal entanglement entropy of the ground states with a quarter of the filling as the function of $V$.
(d)(e)(f) The scaling of the biorthogonal entanglement entropy from (a), the self-norm entanglement entropy from (b) at $V=2$ and $V=4$,
and the biorthogonal entanglement entropy from (c) at $V=2$ and $V=7$.}
\label{figEE}
\end{figure}

\section {Entanglement entropy}
\label{sec:Entanglement}
In the half filling, the extended hard-core BHN can be mapped to the spin-1/2 chain,
\begin{align}
\hat{H} &= \sum_{l} [\frac{t}{2} (\sigma_{l}^{x}\sigma_{l+1}^{x} + \sigma_{l}^{y}\sigma_{l+1}^{y}) + \frac{V}{4} \sigma_{l}^{z}\sigma_{l+1}^{z} \nonumber \\
            & \hspace{1.5cm} +  i \frac{\gamma}{2} (\sigma_{l}^{x}\sigma_{l+1}^{y} - \sigma_{l}^{y}\sigma_{l+1}^{x}) -\frac{V}{2} \sigma_{l}^{z}  + \frac{V}{4}],
            \label{spinBHN}
\end{align}
by performing the transformations between the bosonic operators and the spin operators,
\begin{align}
\sigma_{l}^{+} &= b_{l}, \\
\sigma_{l}^{-}  &= b_{l}^{\dagger}, \\
\sigma_{l}^{z} &= 1 - 2b_{l}^{\dagger}b_{l}, 
\end{align}
where $\sigma_{l}^{+} = (\sigma_{l}^{x} + i\sigma_{l}^{y})/2$, $\sigma_{l}^{-} =(\sigma_{l}^{x} - i\sigma_{l}^{y})/2$ are raising and lowering operators, and $\sigma_{l}^{x},\sigma_{l}^{y},\sigma_{l}^{z}$ are Pauli matrices.

In the case of $\gamma=0$, the model in Eq.(\ref{spinBHN}) is the well-known (Hermitian) XXZ model, which undergoes the Berezinskii-Kosterlitz-Thouless (BKT) transition between the doubly degenerated antiferromagnetic phase and the gapless XY phase. In the bosonic language, 
such two ground states are named as the charge density wave (CDW) phase and the gapless phase.
In order to describe the phase transition, we introduce the entanglement entropy between a part $A$ and a part $B$,
\begin{align}
S_A = -\text{Tr}_{A} (\rho_{A} \ln \rho_{A}),
\end{align}
where $\rho_{A} = \text{Tr}_{B}(\ket{\psi_j}\bra{\psi_j})$, and $\ket{\psi_j}$ is the $j$th eigenstate of the BHN model.
In the following, we choose the subsystems $A=\{ 1, \cdots, L/2 \}$ and $B=\{ L/2+1, \cdots, L \}$.
For a Hermitian system under PBCs, it is shown that the entanglement entropy scales logarithmically as \cite{calabrese2004entanglement},
\begin{align}
S_{L/2} \propto \frac{c}{3} \ln(L) 
\end{align}
at the critical point of a finite-size chain.
The central charge $c=1$ for the XXZ model that can be described by the conformal field theory (CFT) \cite{calabrese2004entanglement}.

\begin{figure}[tb]
\includegraphics[width=8.6cm]{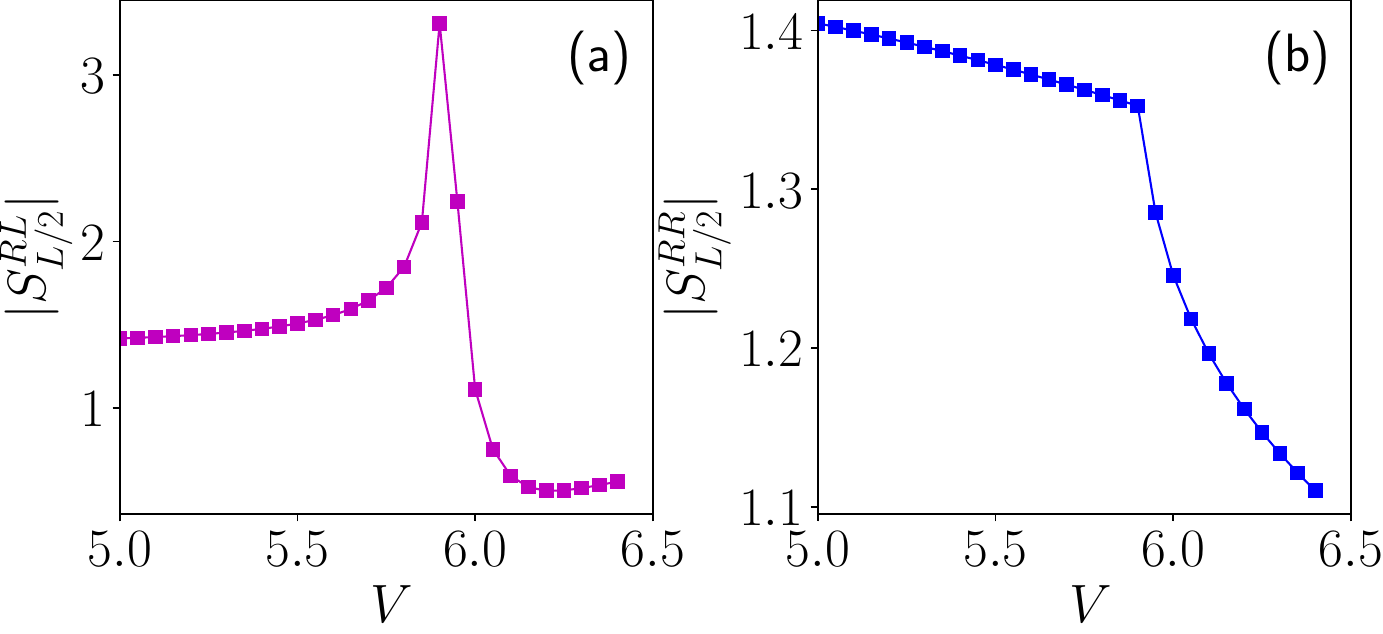}
\caption{ Entanglement entropy of the first-excited states at $L=24$ at $t=1$ and $\gamma=0.6$ in PBCs. 
(a) The biorthogonal entanglement entropy with respect to $V$, (b) The self-norm entanglement entropy as the function of $V$.} 
\label{figOverlap}
\end{figure}

When $\gamma \neq 0$, the model in Eq.(\ref{spinBHN}) is the generalized XXZ model with a complex Dzyaloshinskii-Moriya interaction (DMI), 
which is a non-Hermitian system. 
Although the Hamiltonian in Eq.(\ref{spinBHN}) is non-Hermitian, the ground-state energy (the state with the lowest real part of eigenvalues) is real for any $\gamma$.
To verify whether the entanglement can describe the phase transition of the non-Hermitian XXZ chain, we compute both the biorthogonal entanglement entropy $S_{L/2}^{RL}$ and the self-norm entanglement entropy $S_{L/2}^{RR}$ of the ground state. 
The biorthogonal and self-norm reduced density matrix are defined by the combination of the biorthogonal eigenstates $\rho_{A}^{RL} = \text{Tr}_{B}(\ket{\psi_0^{R}}\bra{\psi_0^{L}})$ and only the right density matrix $\rho_{A}^{RR} = \text{Tr}_{B}(\ket{\psi_0^{R}}\bra{\psi_0^{R}})$, respectively \cite{chang2020entanglement,herviou2019entanglement}. 
The global phase diagram of the non-Hermitian XXZ that is obtained by the sudden change of the first-excited energy from a real value to a complex value [c.f. Fig.\ref{figPhaseDiagram}(c) and (d)] is demonstrated in Fig.\ref{figPhaseDiagram}(a), where we find that the complex DMI enlarges the gapless regime compared to the original Hermitian XXZ model (The critical value $V_c  > 2$ for a nonzero $\gamma$) . 
The CDW phase is characterized by the CDW order parameter $O_{\text{CDW}}= \lim_{r \rightarrow \infty}C_{1, r}$, where $C_{1,r}=(1 - 2n_{1})(1 - 2n_{r})$ [see Fig.\ref{figPhaseDiagram}(b)].
The gapless phase is described by both the biorthogonal and the self-norm entanglement entropy of the ground state [see Fig.\ref{figEE}(a)(b)(d)(e)], 
with a central charge $c=1$ as the Hermitian XXZ model. 

The BHN model shown in Eq.(\ref{eqHNHC}) (or in Eq.(\ref{spinBHN})) has a $\mathcal{PT}$ symmetry, the system thus exists either real eigenenergies or complex eigenenergies in conjugated pairs.
Consequently, the BHN model can in principle exhibit a $\mathcal{PT}$ transition between the $\mathcal{PT}$ symmetric phase and the $\mathcal{PT}$ broken phase.
A $\mathcal{PT}$ transition in the first-excited state has recently been discussed in the Ref.[\onlinecite{zhang2022symmetry}] for the FHN model.
The $\mathcal{PT}$ transition in the first-excited state is confirmed and presented in Fig.\ref{figPhaseDiagram}(a) in the BHN model.
In the following, we will instead investigate the properties of quantum entanglement in the first-excited state for such a $\mathcal{PT}$ transition. 

To achieve it, we shall calculate both the biorthogonal entanglement entropy $S_{L/2}^{RL}$ and self-norm entanglement entropy $S_{L/2}^{RR}$ for the first-excited states.
We note that the first-excited states are doubly degenerated in the $\mathcal{PT}$ broken regime, which seems to bring additional problems to compute the entanglement entropy.
However, thanks to the $\mathcal{PT}$ symmetry, the first-excited states can be distinguished according to the imaginary part of the eigenvalues. 
In the following, we will perform the calculations based on the first-excited state with the negative imaginary part of the energy.
We find that both $S_{L/2}^{RL}$ and $S_{L/2}^{RR}$ exhibit a sudden change when varying the interaction as shown in Fig.\ref{figOverlap}, 
indicating a phase transition happens.
We point out that it is impossible to extract the critical exponents using either $S_{L/2}^{RL}$ or $S_{L/2}^{RR}$. 
The study of exceptional points using the finite-size scaling theory of the entanglement entropy is open. 
Our results indicate that the entanglement entropy can serve as a valid quantity for finding critical points.

Finally, we briefly discuss the properties of the entanglement entropy when the system is away from the half filling.
In the case of the nonhalf filling, the ground phases in the whole regime of $\gamma$ are gapless as the system cannot form a CDW phase.
Consequently, the entanglement entropy of the ground phases should exhibit a logarithmic scaling with the central charge $c=1$.
We verify this argument by the performing scaling of the biorthogonal entanglement entropy as shown in Fig.\ref{figEE}(e)(f). 

\begin{figure}[tb]
\includegraphics[width=9.1cm]{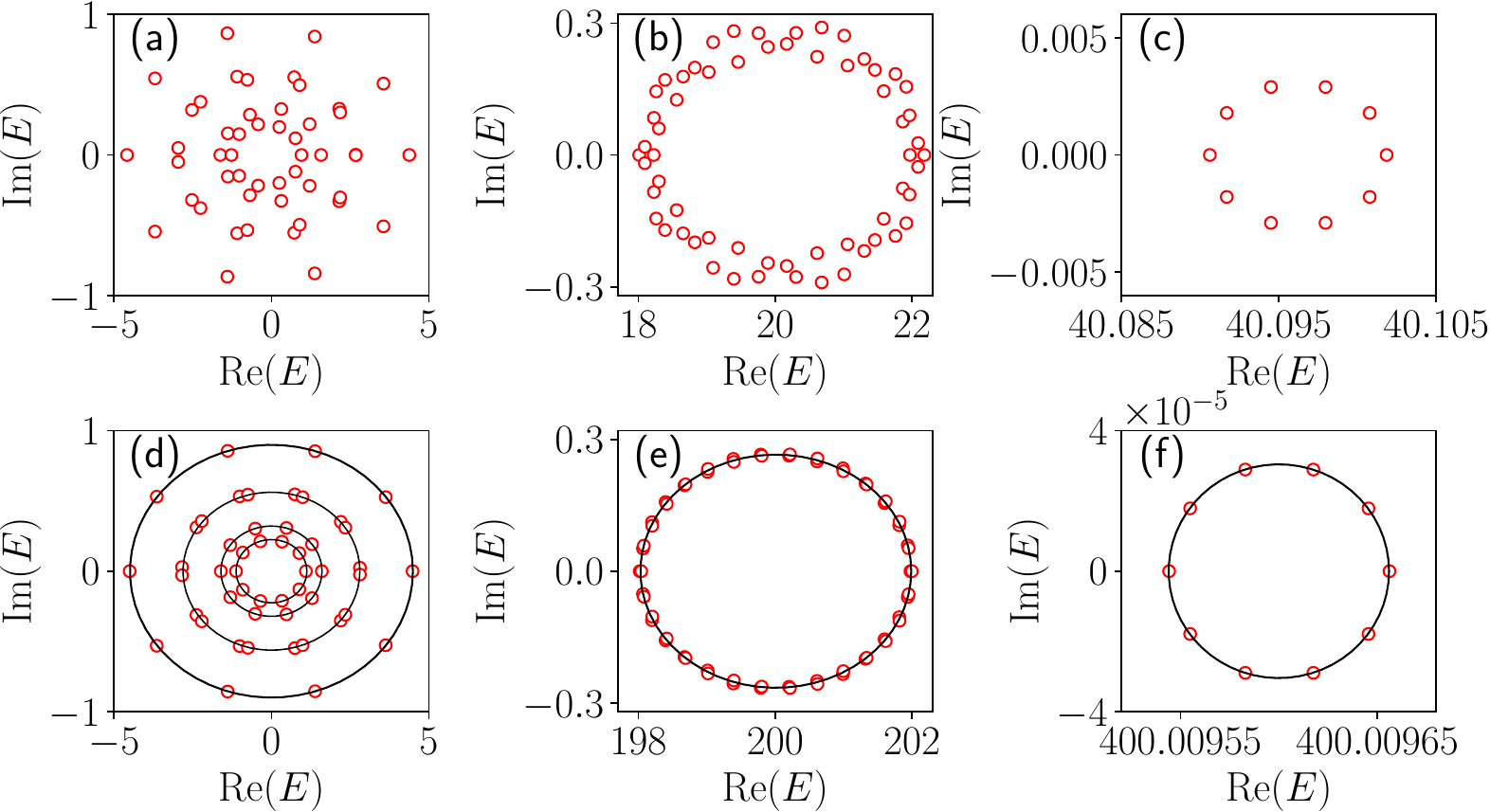}
\caption{The spectral clusters of the BHN model with $L=10$ and $N=3$ at $t=1$ and $\gamma=0.2$.
(a)(b)(c) Spectral clusters for $n_s=1, 2, 3$ and $V=20$, (d)(e)(f) spectral clusters for $n_s=1,2,3$ and $V=200$. }
\label{figCluster}
\end{figure}

\section{Spectral clusters}
\label{sec:EnergySpectrum}
In this section, we shall turn to the energy spectra to investigate the effects of the interaction. 
It is easy to show that the single-particle energy spectrum which can be exactly obtained for the BHN model is a closed curve in the case of noninteractions \cite{zhang2022symmetry,sun2023aufbau}.
However, the spectral structure of the BHN model is too complicated to study under finite interactions. 
Recently, it was found interestingly that the energy spectrum of the FHN model away from the half filling would be separated into spectral clusters [c.f. Fig.\ref{figCluster}(a)(b)(c)] when the interaction strength $V$ is larger than the range of each cluster \cite{zhang2022symmetry}. 
It is worth while to note that energy spectra can also form clusters even in the half filling, which is not discussed in Ref.\cite{zhang2022symmetry}.
The spectral clusters in the half filling would be located only in the real-energy axis as energies are always real under strong interactions [c.f. Fig.\ref{figPhaseDiagram}(c)(d)].
In this following, we will study the BHN model for arbitrary filling under strong interactions to investigate universal properties of energy spectral structures.

\begin{figure}[tb]
\includegraphics[width=8.3cm]{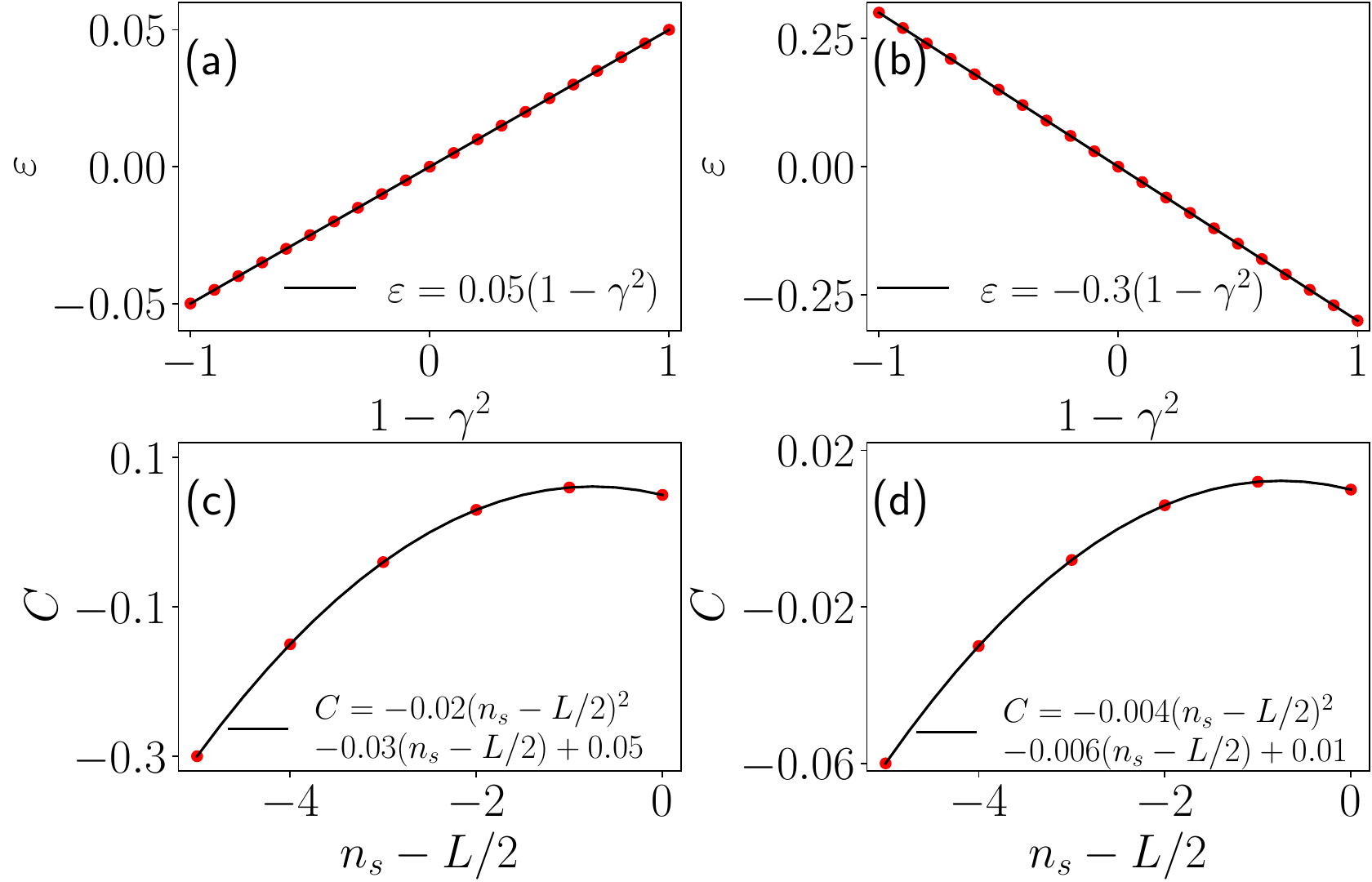}
\caption{ The energy shift in the half-filled BHN model with $L=12$ at $t=1$ in PBCs.
(a)(b) The energy shift $\varepsilon$ for the cluster $n_s = N$ and the $n_s = 1$ with respect to $\gamma$ at $V=40$, respectively.
(c)(d) The coefficients $C$ in $\varepsilon$ as the function of $n_s$ for $V=40$ and $V=200$ at $\gamma=0.2$, respectively.}
\label{figmodifiedEE}
\end{figure}

It can be seen from the Fig.\ref{figCluster}(a)(b)(c) that the energy spectrum is symmetrically distributed in the complex plane \cite{zhang2022symmetry}.
Assume the system has $N$ particle numbers, the energy spectrum would form $N$ clusters. 
The central position of each cluster $E_c$ is round $(n_s-1)V$.
Here, we use $n_s$, with $n_s =1,\cdots,N$, to label these clusters individually. As the system has a particle-hole symmetry, 
the properties of clusters for $L-N$ particles are the same as those for $N$ particles.
Let us first study the properties of the central positions $E_c$ of these energy clusters. We show that the central positions of these clusters are not exactly 
located at $(n_s-1)V$. Instead, they are \cite{zhang2022symmetry},
\begin{equation}
E_c=(n_s-1)V+\varepsilon,
\label{equ:E_c}
\end{equation}
with,
\begin{equation}
\varepsilon \approx C(t^2-\gamma^2).
\label{equ:epsilon}
\end{equation}
It was shown that $C=2/V$ for the cluster $n_s=N$ in Ref.\cite{zhang2022symmetry}.
However, we argue that this coefficient is not always valid for other clusters. 
For example, we find that $C=2/V$ for the cluster $n_s=N$, but $C=-L/V$ for the cluster $n_s=1$ in half filling [c.f. Fig.\ref{figmodifiedEE}(a)(b)].  
Moreover, we find that $C$ is indeed a quadratic function with respect to $n_s$ for other cases.  
The coefficient $C$ and the relation $n_s-N$ is found to satisfy the following quadratic function, 
\begin{equation}
C=-\frac{4}{V(L/2-1)}(n_s-L/2)^2 - \frac{2(L/2-3)}{V(L/2-1)}(n_s-L/2) + \frac{2}{V},
\label{eqEnergyCoeff}
\end{equation}
in half filling. This argument is verified by numerical simulation using $ L=12$ at $V=40$ and $V=200$ [c.f. Fig.\ref{figmodifiedEE}(c)(d)].
For instance, we find that $C=0.05 = 2/V$ when $n_s=N=L/2$, and $C=-0.3 = -L/V$ when $n_s=1$ at $V=40$, which are consistent with Eq.(\ref{eqEnergyCoeff}). 

Next, we turn to study the shapes of clusters. 
Interestingly, we find that all spectral clusters would form perfect ellipses if the interaction $V$ is strongly enough [c.f. Fig.\ref{figCluster}(d)(e)(f)]. 
Let us first discuss the rightmost cluster ($n_s=N$) which is the simplest case as shown in Fig.\ref{figCluster}(f).
We fit the data in Fig.\ref{figCluster}(f), finding that the function is an elliptic curve,
\begin{equation}
\frac{(x-x_0)^2}{a^2}+\frac{y^2}{b^2}=1,
\end{equation}
with the fitting parameters are $x_0\approx 400.0096$, $a\approx 5.6001\times 10^{-5}$, and $b\approx 3.0401\times 10^{-5}$.
Here, $2a$ and $2b$ are the lengths of the major and minor axes of the ellipse, 
$x_0=E_c$ is the position of the centre of the ellipse, which satisfies $x_0 \approx (n_s-1)V+2(t^2-\gamma^2)/V$.

\begin{figure}[tb]
\includegraphics[width=8.6cm]{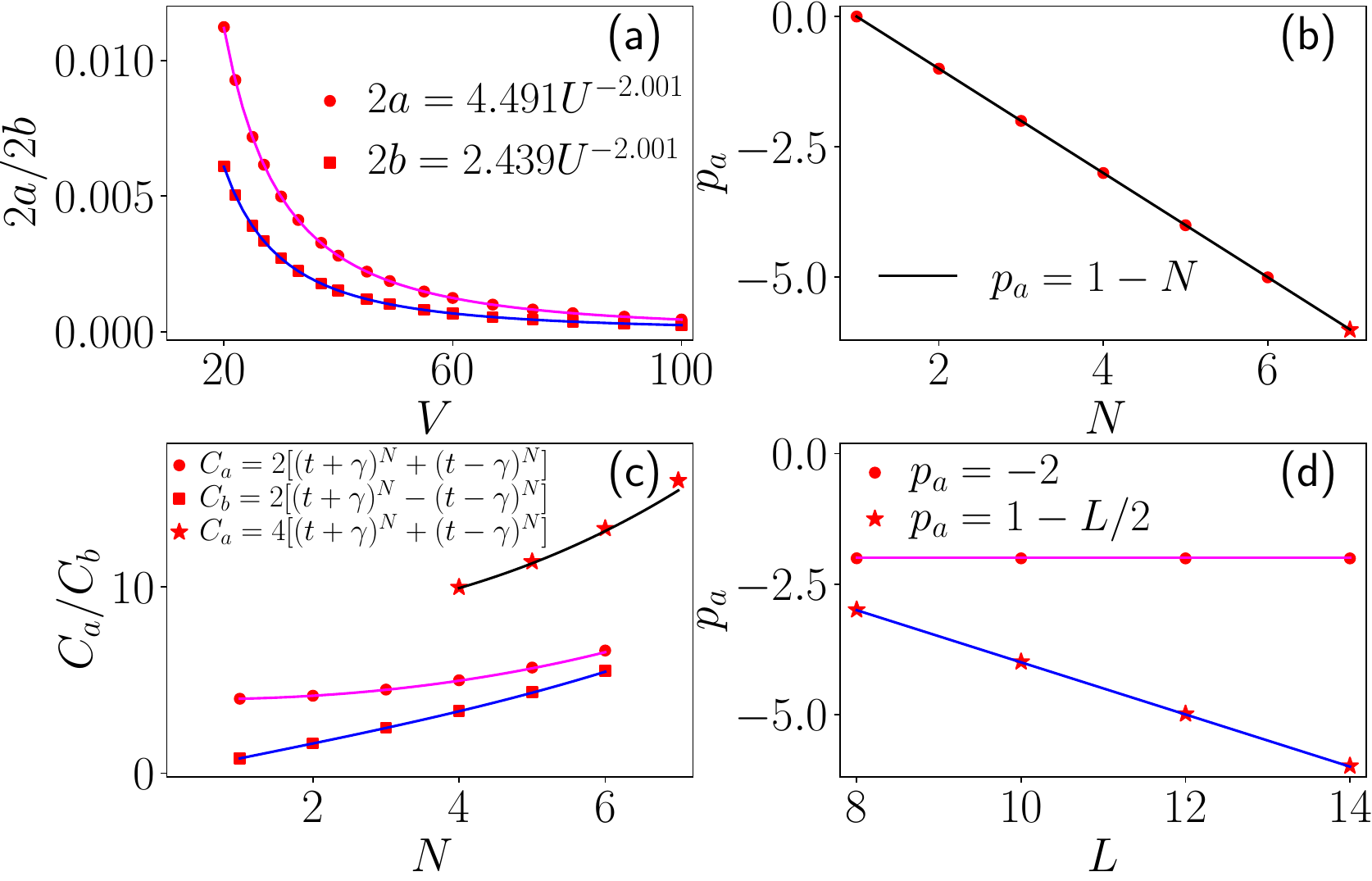}
\caption{ The scaling parameters of the major and minor axes at $t=1$ and $\gamma=0.2$.
(a) The scaling of major and minor axes of the ellipse for the cluster $n_s=3$ in the BHN model with $L=10$ and $N=3$ particles.
(b)(c) The exponent $p_a$ and coefficients $C_a$ and $C_b$ with respect to $N$ for cluster $n_s=N$ up to $L=14$,
(d) The exponent $p_a$ as the function $L$ for cluster $n_s=1$. The star symbols in (a)(b)(c) denote the data in the half filling $N=L/2$.}
\label{figCylinder}
\end{figure}

We are surprised to find that the major axis $2a$ and minor axis $2b$ obey a universal scaling law [c.f. Fig.\ref{figCylinder}], which is given by, 
\begin{align} 
2a &= C_aV^{p_a} +V_{0a}, \label{equ:ab_milva} \\
2b &= C_bV^{p_b} +V_{0b},
\label{equ:ab_milvb}
\end{align}
where $C_a$, $C_b$, $p_a$, $p_b$, $V_{0a}$ and $V_{0b}$ are fitting parameters.
To obtain the accurate values of $C_a$, $C_b$, $p_a$ and $p_b$, we fit the major and minor axes $2a$ and $2b$ with respect to $V$ of the rightmost cluster 
for $L=8,10,12,14$ and $n_s=N$. 
 The results are shown in Fig.\ref{figCylinder}(a)(b)(c) and Table \ref{tab:ab_U_L10_14} [see Appendix \ref{App:A} for details], 
where we find that when the system is away from the half filling, $C_a$, $C_b$, $p_a$ and $p_b$ are given as,
\begin{align} 
	C_a = 2[(t+\gamma)^N+(t-\gamma)^N], ~~~ p_a=1-N, \label{equ:ab_milv1} \\
	C_b = 2[(t+\gamma)^N-(t-\gamma)^N], ~~~ p_b=1-N.
	\label{equ:ab_milv2}
\end{align}
Above scaling exponents can be obtained from the perturbation theory \cite{zhang2022symmetry}.
When the system is in the half filling, the eigenenergies are alway real for large interactions \cite{zhang2022symmetry}, 
thus only $C_a$ and $p_a$ need to be derived, which are given by, 
\begin{equation}
	\begin{aligned} 
		C_a = 4[(t+\gamma)^N+(t-\gamma)^N], ~~~ p_a=1-L/2.
		\label{equ:ab_milv3}
	\end{aligned}
\end{equation}
We note that $V_{0a}=V_{0b}=0$ for an arbitrary filling.

To verify whether above scaling exponents $p_a, p_b$ obtained for $n_s=N$ are also valid for other clusters, we study the spectral properties of the first cluster ($n_s=1$).
It can be seen from Fig.\ref{figCluster}(d), the energy spectrum forms parallel ellipses for the first cluster.
The major axis $2a$ and minor axis $2b$ of the outermost ellipse are numerically fitted by using the Eq.(\ref{equ:ab_milva}) and Eq.(\ref{equ:ab_milvb}).
The parameters $C_a$, $C_b$, $p_a$, $p_b$, $V_{0a}$ and $V_{0b}$ are shown in Table \ref{tableB} [see Appendix \ref{App:A} for details], where we find that $p_a$ and $p_b$ remain,
\begin{equation}
p_a=p_b=1-L/2,
\end{equation}
for the half-filled system.  While if the system is away from the half filling, $p_a$ and $p_b$ satisfy,
\begin{equation}
p_a = p_b = -2.
\end{equation}
for all $N$ and $n_s$, except for the single-particle case ($N=1$), where the interaction is useless [c.f. Fig.\ref{figCylinder}(d)]. 

For more general cases, that is $2 \leq n_s \leq N-1$, it seems that the exponents $p_a$ and $p_b$ satisfy,
\begin{equation} 
p_a=p_b=1-n_s.
\label{eq:ab_arbitrary}
\end{equation}
for arbitrary fillings. 
However, we have to note that there are several exceptions where $p_a$ and $p_b$ are not consistent with this scaling rule in Eq.(\ref{eq:ab_arbitrary}).
The reason for this may be that the clusters for $2 \leq n_s \leq N-1$ have complex structures [see Fig.\ref{figCluster}(b)(e)], one has to increase the interaction $V$ to have a perfect ellipse.
However, when $V$ is too large, the fitting functions $V^{p_a}$ and $V^{p_b}$ would become exponential small, which we think may lead to the instability of the fitting.
We note that other possible relations between $p_a, p_b$ and $n_s$ cannot be ruled out. This problem is left for future study.

Finally, we investigate the numbers of energy levels on a single elliptic ring.
It is found that the energy levels distribute symmetrically in the elliptic lines, in which the number of the levels $N_q$ is related to the chain length $L$.
Strictly speaking, we find that the number of levels $N_q$ on the outermost elliptic ring is, 
\begin{equation}
N_q = L,
\end{equation}
for the clusters $n_s=1$ and $n_s=N$  [see Fig.\ref{figCluster}(d)(f)]. 
For other clusters $ 2 \leq n_s \leq N-1$, the number of energy levels $N_q$ is given by,
\begin{equation} 
N_q=(n_s+L/2-N-1)L,
\end{equation}
dependent on $L$, $N$ and $n_s$. 
For example, $N_q=30$ for $L=10$, $N=3$ and $n_s=2$ as shown in Fig.\ref{figCluster}(e). Besides,
the energy values $E$ on the outermost ellipse can be derived as,
\begin{equation}
	\begin{aligned} 
		E(q)= E_c + a \cos(q)+ib \sin(q),
		\label{equ:ab_milv5}
	\end{aligned}
\end{equation}
with $q=2n\pi/N_q$, and $n=0,1,\cdots, N_q -1$. The energy spectrum in Eq.(\ref{equ:ab_milv5}) is similar to the single-particle energy spectrum, indicating that one may analyze the many-body physics using the concepts or techniques from the single-particle picture.
In addition, universal properties of many-body systems under intermediate interactions is completely unknown so far, which is beyond the scope of this work and left for future research.

\section {Conclusion}
\label{sec:conclusion}
We have studied the entanglement entropy and spectral clusters of the extended hard-core BHN model in PBCs.
We show that the extended hard-core BHN model undergoes a phase transition in the ground state that is accompanied by a $\mathcal{PT}$ transition in the first excited state discussed in FHN model \cite{zhang2022symmetry}. The phase transition can be described by the CDW order parameter of the ground state, the biorthogonal and self-norm entanglement entropy of the first-excited state.
  
Furthermore, we study the properties of the energy eigenvalues of each clusters led by the nearest-neighbor interaction, 
showing that the shape of each cluster is an elliptic function when the  interaction is strongly enough.
We explore the universal scaling laws for the size of the ellipses with respect to the interaction and derive the corresponding scaling exponents.
Finally, we analyze the number of energy levels of each ellipse and find a universal express for the number of levels in the outermost circles of clusters.

\begin{acknowledgments}
This work is supported by "the Fundamental Research Funds for the Central Universities, NO. NS2023055"
and is partially supported by the High Performance Computing Platform of Nanjing University of Aeronautics and Astronautics.

\end{acknowledgments}

\bibliographystyle{apsrev4-2}
\bibliography{refHNcluster}

\clearpage
\begin{widetext}
\appendix
\section{Numerical data for $n_s=N$ and $n_s=1$}
\renewcommand{\thetable}{\Alph{section}\arabic{table}}
\setcounter{table}{0}
\label{App:A}
In this appendix, we display the data for $n_s=N$ in Table \ref{tab:ab_U_L10_14} and the data for $n_s=1$ in Table \ref{tableB}, respectively.
Please see below for details.
\begin{table}[h]
	\caption{The fitting parameters of the major and minor axes $2a$ and $2b$ with respect to $V$ from $V=40$ to $V=100$ with the step $dV=5$ for the cluster $n_s=N$ with $L=8,10,12,14$, $t=1$ and $\gamma=0.2$. We note that $V=20$ to $V=60$ with the step $dV=5$ is used for $L=14$, $N=7$ to fit as $V^{p_a}$ is too small when $V$ is large.
	\label{tab:ab_U_L10_14}}
	\begin{tabular}{|ccccccc|}
		\hline
		$L$ & $N$ & $1-N$ & $C_a$ & $p_a$ & $C_b$ & $p_b$\\
		\hline
		 8& 1& -0& 4.000& -0.000& 0.800& -0.000\\
		 8& 2& -1& 4.160& -1.000& 1.600& -1.000\\
		 8& 3& -2& 4.505& -2.001& 2.427& -2.000\\
		 8& 4& -3& 9.981& -3.001&     - &      - \\
		10& 1& -0& 4.000& -0.000& 0.800& -0.000\\
		10& 2& -1& 4.160& -1.000& 1.600& -1.000\\
		10& 3& -2& 4.491& -2.001& 2.438& -2.001\\
		10& 4& -3& 5.007& -3.002& 3.334& -3.001\\
		10& 5& -4& 11.35& -4.002&     - &      - \\
		12& 1& -0& 4.000& -0.000& 0.800& -0.000\\
		12& 2& -1& 4.160& -1.000& 1.600& -1.000\\
		12& 3& -2& 4.491& -2.001& 2.438& -2.001\\
		12& 4& -3& 4.991& -3.001& 3.344& -3.001\\
		12& 5& -4& 5.692& -4.002& 4.335& -4.001\\
		12& 6& -5& 13.13& -5.002&     - &      - \\
		14& 1& -0& 4.000& -0.000& 0.807& -0.000\\
		14& 2& -1& 4.160& -1.000& 1.600& -1.000\\
		14& 3& -2& 4.491& -2.001& 2.438& -2.001\\
		14& 4& -3& 5.991& -3.001& 3.344& -3.001\\
		14& 5& -4& 5.673& -4.002& 4.353& -4.002\\
		14& 6& -5& 6.587& -5.003& 5.507& -5.002\\
		14& 7& -6& 15.70& -6.008&     - &      - \\
		\hline
	\end{tabular}
\end{table}

\begin{table}[h]
	\caption{The fitting parameters of the major and minor axes $2a$ and $2b$ with respect to $V$ from $V=40$ to $V=100$ with the step $dV=5$ for the cluster $n_s=1$ with $L=8,10,12,14$, $t=1$ and $\gamma=0.2$.}
	\begin{tabular}{|ccccccccc|}
		\hline
		$L$ & $ N $&$1-N$& $C_a$&$p_a$&$ V_{0a} $&$C_b$&$p_b$&$ V_{0b} $\\
		\hline
		 8& 1& -0&  0.000& -0.000& 4.000&  0.0000&  0.000& 0.800\\
		 8& 2& -1& -3.539& -2.005& 6.928&  0.0277& -2.000& 1.386\\
		 8& 3& -2& -11.07& -1.989& 6.472& -2.9880& -2.002& 1.294\\
		 8& 4& -3& 95.66& -2.992& 0.000& -& -& -\\
	        10& 1& -0&  0.000&  0.000& 4.000& 0.000& 0.000& 0.800\\
		10& 2& -1& -1.527& -2.006& 7.391& 0.0119& -2.000& 1.478\\
		10& 3& -2& -8.934&  -1.999& 8.988&-0.5884& -2.001& 1.798\\
		10& 4& -3&-11.56& -1.999& 6.928&  -5.605& -2.007& 1.386\\
		10& 5& -4& 376.57& -3.990& 0.000&  -& -& -\\
		12& 1& -0& -0.000 & -0.000& 4.000&  0.000& -0.000& 0.800\\
                 12& 2& -1&-0.7764 & -2.007& 7.608&  0.006& -2.000& 1.522\\
	         12& 3& -2& -5.260 & -2.003& 10.13&  -0.1479& -2.000& 2.026\\
		12& 4& -3& -13.84 & -1.996& 10.45&  -1.995& -2.006& 2.091\\
		12& 5& -4& -11.05 & -2.000& 7.208&  -7.248& -1.999& 1.442\\
		12& 6& -5& 1551.37 & -4.989& 0.000&  -& -& -\\
		14& 1&  -0& -0.000 & -0.000& 4.000&  -0.000& -0.000& 0.800\\
		14& 2&  -1& -0.4431 & -2.007& 7.727&  0.00345& -2.000& 1.545\\
		14& 3&  -2& -3.137 & -2.005& 10.73&  -0.0434& -1.999& 2.146\\
		14& 4& -3& -10.23 & -2.000& 12.31&  -0.706& -2.003& 2.462\\
		14& 5&  -4& -17.46 & -1.995& 11.52&  -3.876& -2.005& 2.304\\
		14& 6&  -5& -10.41 & -2.001& 7.391&  -8.433& -1.993& 1.478\\
		14& 7&  -6& 6591.26 & -5.987& 0.000&  -& -& -\\
		\hline
	\end{tabular}
	\label{tableB}
\end{table}

\end{widetext}

\end{document}